\documentclass[aps,prl,reprint,superscriptaddress,showpacs]{revtex4-1}
\usepackage{graphicx}
\usepackage{subfigure}
\usepackage{dcolumn}
\usepackage{bm}
\usepackage{hyperref}
\usepackage{indentfirst}
\usepackage{braket}
\usepackage{amsmath}
\usepackage{amssymb}
\hypersetup{colorlinks=true, citecolor=blue, urlcolor=blue, linkcolor=blue}
\begin{document}


\title{Charge Density Wave in a Doped Kondo Chain}


\author{Yixuan Huang}
\affiliation{Texas Center for Superconductivity, University of Houston, Houston, Texas 77204, USA.}
\author{D. N. Sheng}
\affiliation{Department of Physics and Astronomy, California State University, Northridge, California 91330, USA.}
\author{C. S. Ting}
\email{ting@uh.edu}
\affiliation{Texas Center for Superconductivity, University of Houston, Houston, Texas 77204, USA.}


\date{\today}

\begin{abstract}
We report the existence of the charge density wave (CDW) in the ground state of 1D Kondo lattice model at the filling of n=0.75 in the weak coupling region. The CDW is driven by the effective Coulomb repulsion mediated by the localized spins. Based on our numerical results using the density matrix renormalization group method, we show that the CDW phase appears in the paramagnetic region previously known as the Tomonaga-Luttinger liquid. The emergence of this phase serves as an example of CDW phase induced without bare repulsive interactions, and enriches the phase diagram of the 1D Kondo lattice model.
\end{abstract}


\maketitle

\textit{Introduction.---}
As a fundamental model for strongly correlated system describing heavy fermion materials\cite{hossain2005coexistence,ochiai1990heavy} and quantum magnetism\cite{RevModPhys.73.797}, the Kondo lattice (KL) model\cite{doniach1977kondo,PhysRevB.20.1969,fazekas1991magnetic} has been intensively studied over the last three decades, especially in one dimension\cite{tsunetsugu1993phase,tsunetsugu1997ground}. While most studies have focused on the Kondo effect and the Ruderman- Kittel-Kasuya-Yosida (RKKY)\cite{ruderman1954indirect,kasuya1956theory,yosida1957magnetic} interaction, the effective Coulomb repulsion mediated by the localized spins was neglected. In the 1D bosonization theory, if the interaction is strong enough, the system undergoes a phase transition from the Tomonaga-Luttinger liquid (TLL) to an insulating phase. The TLL phase with a large Fermi surface has been found in the weak coupling\cite{shibata1996friedel,shibata1997one,khait2018doped} of 1D KL model. Whether the effective Coulomb repulsion is strong enough to induce a charge ordered phase at commensurate filling remains to be an open question. 

The origin of the effective repulsive interaction in a KL model is proposed by a strong-coupling perturbation expansions\cite{hirsch1984strong}. However, for 1D the strong coupling region is dominant by the Kondo effect which results in ferromagnetism\cite{mcculloch2002localized,peters2012ferromagnetic} at less than half filling. At half filling, the insulating phase is caused by the formation of the Kondo singlet and the Coulomb repulsion is suppressed. The 1D KL model at quarter filling has also been investigated \cite{xavier2003dimerization,PhysRevB.78.144406} to realize a dimerization of the localized spins induced by RKKY interaction, although there are some controversy about the existence of true dimer order\cite{hotta2006absence,shibata2011boundary}. Thus a charge order may be expected in the next order commensurate filling of $\frac{1}{8}$ or $\frac{3}{8}$. 

The charge order in KL model has been investigated in higher dimensions. Using the dynamical mean field theory (DMFT) method and variational Monte Carlo method, a charge density wave in the weak coupling has been found in both two dimensions\cite{misawa2013charge, motome2010partial,sato2018quantum} and infinite dimensions\cite{peters2013charge} at quarter fillings. The intriguing question remains whether the charge order exists in one dimension, as the DMFT method generally gets less accurate in low dimensions, especially in one dimension. Additionally, the charge order in higher dimensions is stabilized by the Kondo singlet formation, which is different from the 1D case.

\begin{figure}
\centering
\includegraphics[width=0.8\columnwidth]{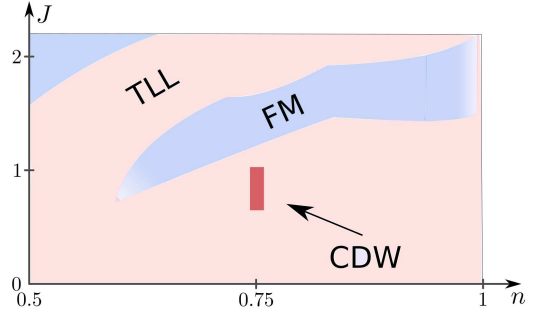}
\caption{\label{Fig1}(Color online) The schematic diagram of the 1D Kondo lattice model in the region of $n>0.5$ and $J<2.2$ where $n$ is the electron density, and $J$ is the coupling between electrons and localized spins. }
\end{figure}

In this letter, we identify a charge density wave (CDW) phase at commensurate filling of $\frac{3}{8}$, corresponding to $n=\frac{3}{4}$ in the 1D KL model, as shown in Fig.\ref{Fig1}. The phase is characterized by a finite density oscillation in arbitrarily long chain, and a vanishing Luttinger parameter. The formation of the CDW opens a gap in the charge sector, similar to the metal-insulator transition in the 1D extended Hubbard model\cite{jeckelmann2002ground,sandvik2004ground}, while the spin part remains gapless. Our results showing the charge ordered phase in 1D at higher order commensurate filling provide an insight to the exotic phase of KL model, and a possible mechanism of the charge order in 1D organic compounds\cite{seo2004toward,monceau2012electronic}, especially the Peierls instabilities of $\left ( Per \right )_{2}  \left [ M \left ( mnt \right )_{2} \right ]$ (M = Pd, Ni and Pt)\cite{pouget2017peierls}. In the existence of a CDW, the behavior of the localized spin is the result of the competition between RKKY interaction and CDW, which provides an example of how the CDW interacts with the localized spins.

\textit{Model and Methods.---}
In this work, we consider the standard one-dimensional Kondo Lattice model which describes the itinerant electrons coupled to the localized spins on every unit with a SU(2) symmetric antiferromagnetic interaction:

\begin{equation}
\label{eq1}
\mathcal{H}=-t{\displaystyle {\displaystyle {\textstyle {\displaystyle \sum_{i=1,\sigma}^{L-1}c_{i,\sigma}^{\dagger}}c_{i+1,\sigma}+H.c.}}+J\sum_{i=1,\sigma}^{L}\overrightarrow{S_{i}}\cdot\overrightarrow{s_{i}}}
\end{equation} 

The first term in the Hamiltonian is the hopping term where $c_{i,\sigma}^{\dagger}$ refers to the creation operator of an electron on site $i$ with spin index $\sigma$, the second term describes the spin-spin interaction where $\overrightarrow{S_{i}}$ denotes the localized spin-$\frac{1}{2}$ and $\overrightarrow{s_{i}}=\frac{1}{2}\sum_{\alpha,\beta}c_{i,\alpha}^{\dagger}\overrightarrow{\sigma}_{\alpha,\beta}c_{i,\beta}$ (with Pauli matrices $\overrightarrow{\sigma}$) the conduction electron spin.

The strong correlations in the weak coupling between electrons and localized spins makes it notoriously difficult for exact solutions. Thus numerical method becomes important in order to determine the phase diagram.

We use the U(1) Density Matrix Renormalization Group\cite{PhysRevLett.69.2863,PhysRevB.48.10345,schollwock2011density} (DMRG) method with open boundary condition for lattice size up to $L=208$. Calculations are performed using the ITensor library\cite{ITensor}. Smaller sizes are also used for finite size extrapolation. The largest bond dimension is 7,000 during the sweeps. 140 sweeps with increasing bond dimension were used in order to reach stable and convergent ground state. The cutoff error during the last few sweeps is $10^{-7}$. The hopping parameter $t$ and the lattice spacing is set to unity to fix the energy scale.

\begin{figure}
\centering
\input{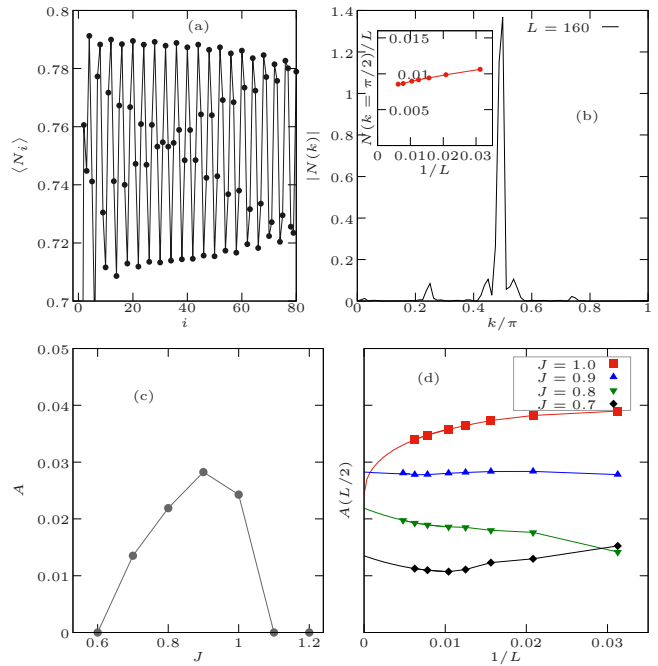}
\caption{\label{Fig2}(Color online) $(a)$ is the electron density at $J=0.9,L=160,n=\frac{3}{4}$. Only half of the lattice is shown. $(b)$ is fourier transform of the density oscillation in $(a)$. The dominant peak shows up at $k=\frac{\pi}{2}$. The inset of $(b)$ shows the intensity of the $k=\frac{\pi}{2}$ peak divided by the length. $(c)$ is the $J$ dependence of the oscillation amplitude in the middle after finite size extrapolation. The extrapolation in $(d)$ is conducted in this way. If the fitting agrees well with the power decay of the Friedel oscillation\cite{shibata1996friedel} in TLL then the order parameter is considered zero when $L\rightarrow \infty $, otherwise we use a least-square fit to the second order of polynomials in $1/L$. For $J$ close to the transition we find it better to fit to $1/\sqrt{L}$ as the similar scaling in TLL.}
\end{figure}

\textit{CDW order at $n=\frac{3}{4}$.---}
Under the open boundary condition, the electron density in the CDW phase shows modulation in real space as a result of spontaneous symmetry breaking. While the Mermin-Wagner theorem\cite{PhysRevLett.17.1133} forbids any spontaneous breaking of  continuous symmetry in 1D, CDW only breaks the lattice translational symmetry which is discrete. In Fig.\ref{Fig2}$(a)$ we show $\left \langle N_{i} \right \rangle$ at $J=0.9,L=160$. There are strong oscillations around the average electron density with the amplitude $\approx  0.03$. The oscillation decays very slowly away from the boundary and remains finite in the middle of the chain. This allows us to define the order parameter as $A=\lim_{L\rightarrow \infty }A\left( L/2\right) $ the amplitude of electron density oscillation in the middle. 

The finite size extrapolation is needed to determine the order parameter in the large system limit. The extrapolated order parameter is plotted against $J$ in Fig.\ref{Fig2}$(c)$. When $J$ is close to the critical point slightly above 1.0, the order parameter quickly rises from 0, and then decreases slowly to 0 as $J$ decreases. Fig.\ref{Fig2}$(d)$ shows the finite size extrapolation of the order parameter for various coupling. Inside the parameter range from $J=0.7$ to $1.0$ $A\left( L/2\right)$ has a weak dependence of $L$, and remains a finite value as $L$ goes to infinity. 

The Fourier transform of the electron density in Fig.\ref{Fig2}$(a)$ is plotted in Fig.\ref{Fig2}$(b)$. We have used the smoothed Fourier transform in order to minimize the effect caused by the open boundary. The details of the window function that we use are discussed in Refs.\cite{vekic1993smooth,white2002friedel}. Here the CDW phase is dominant by a single peak at $k=\frac{\pi }{2}$, which corresponds to the oscillation period of 4 lattice spacing. There may be a superposition of other oscillation frequency such as $k=\frac{\pi }{4}$, but they all vanish in the thermodynamic limit as illustrated later. The inset of Fig.\ref{Fig2}$(b)$ shows the intensity of the dominant peak divided by the lattice size. The intensity has a almost linear dependence of L and remains finite after the extrapolation. 

The structure factor of the charge ordered phase always scales as the order parameter multiplied by the lattice length. The order parameter defined by the structure factor is essentially equivalent to the definition of the oscillation amplitude in the middle of an infinite chain. In the CDW phase the density oscillation of the infinite chain should be uniform. In that case the Fourier transform of the oscillation is proportional to the oscillation amplitude multiplied by the length, and the two definitions of order parameter only differ by a factor of 2.

The Luttinger parameter $k_{L}$ in the TLL of the 1D KL model has a monotonic decrease as $J$ decreases\cite{shibata1997one,khait2018doped}, indicating a strong repulsive interaction between electrons in the weak coupling region. This could explain the formation of a CDW under strong repulsion. However, in the limit of $J\rightarrow 0$, the system goes back to free 1D electrons with $k_{L}=1$. Thus it is natural to see a critical point where the effective repulsive interaction isn't strong enough to stabilize CDW, and the system goes back to TLL. In the bosonization picture, the Umklapp type scattering term, which carries a fast oscillation phase factor, only appears in the low energy effective Hamiltonian at special fillings\cite{giamarchi2004quantum}.

\begin{equation}
\label{eq4}
\Psi _{R,\uparrow }^{\dagger }\Psi _{R,\downarrow }^{\dagger }\Psi
_{L,\uparrow }\Psi _{L,\downarrow }\propto e^{-i4K_{F}x}e^{i2\sqrt{2}\phi
_{\rho }\left( x\right) }
\end{equation} 

If $4K_{F}=2\pi$, corresponding to $n=1$, the fast oscillation term becomes constant in Eq.\eqref{eq4} and enters in the low energy effective Hamiltonian. If $k_{L}<1$ then this term is relevant in the renormalization procedure, and opens a gap in the charge sector. Higher order terms of Umklapp scattering could also occur at other commensurate fillings. In our case of $n=\frac{3}{4}$, the fourth power term of the Umklapp scattering can occur in the extremely small value of $k_{L}<\frac{1}{16}$. In fact, as we vary $J$ at $n=\frac{3}{4}$, $k_{L}\rightarrow \frac{1}{16}$ when the system is close to the phase transition.

\begin{figure}
\centering
\input{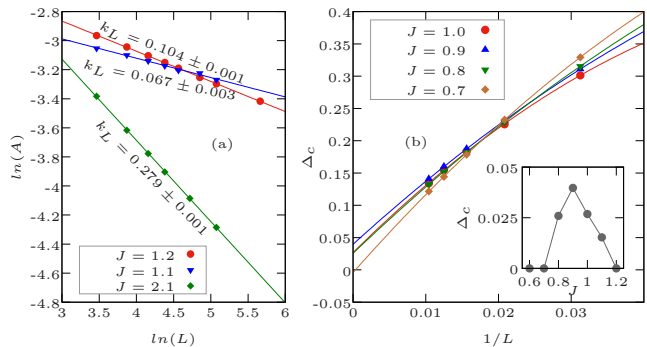}
\caption{\label{Fig3}(Color online) $(a)$ is the log-log plot of the Friedel oscillation amplitude in the middle of various lengths. The Luttinger parameter $k_{L}$ is determined by the $slope =-2k_{L}$. The error of $k_{L}$ is given by the standard deviation of the least square fit. The Luttinger parameter decreases as we lower $J$ into the weak coupling region, which agrees with previous papers. For $J=1$ it cannot be linearly fit, indicating the onset of CDW. $(b)$ is the finite size extrapolation of the charge gap at $n=\frac{3}{4}$. Four examples of $J$ are given for the least-square fit. The inset of $(b)$ shows the extrapolated charge gap for various $J$.}
\end{figure}

In order to determine the Luttinger parameter in the TLL phase, we compare the Friedel oscillation\cite{shibata1996friedel} amplitude at the center for different lattice sizes. The methods that we used to determine the Luttinger parameter are adopted from Ref\cite{white2002friedel}. As shown in Fig.\ref{Fig3}$(a)$, we fit the log-log plot of the oscillation amplitude and find that for $J=1.1$ the Luttinger parameter is 0.067, which is very close to the critical value of $\frac{1}{16}$. In this case, the oscillation decay rate is so slow that we cannot be sure whether or not the oscillation will vanish in the limit of infinite long chain\cite{white2002friedel}. Although it is hard to determine the exact critical value of J, the Luttinger parameter close to the transition point provides another evidence of the emergence of the CDW.

Now we investigate the charge gap in the CDW. Fig.\ref{Fig3}$(b)$ shows the $J$ dependence of the Charge gap at $n=\frac{3}{4}$, which is defined as

\begin{equation}
\label{eq3}
\begin{split}
\Delta _{c}=lim_{L\rightarrow \infty } [ E_{0}\left ( N_{e}= N+2 \right )+E_{0} ( N_{e}=\\ 
N-2 )-2E_{0}\left ( N_{e}= N \right )  ]
\end{split}
\end{equation}

where the $E_{0}\left ( N_{e} \right )$ refers to the ground state energy of a given electron number. Here we set $N=\frac{3}{4}L$ to fix the electron density. We choose $J$ carefully to avoid the ferromagnetic region so that the ground state is always in the $S_{z}^{total}=0$ subspace. The charge gaps here only depend on the ground state energy calculated by the DMRG method, thus are very reliable. In Fig.\ref{Fig3}$(b)$, the charge gap is extrapolated by a least-square fit to the second order of polynomials in $1/L$. A non-zero value of the gap can be distinguished in the thermodynamic limit. The inset of Fig.\ref{Fig3}$(b)$ shows the extrapolated result of the charge gap at different values of $J$. The gap reaches a maximum value at around $J = 0.9$, and has a monotonic decreasing to 0 apart from the peak. Generally in a gapped phase the correlation function either has an exponential decay or decays to a constant at large separations. The electron density oscillation agrees with the latter case here. The emergence of a non-zero charge gap is consistent with the CDW, and together with the CDW order parameter, we can established the TLL-CDW phase boundary at $J\approx 1.1$ and $0.7$. 

The measurement of the Zeeman field needed to close the charge gap could be used as a scale of finite transition temperature $T_{c}\sim \frac{\mu_{B} h_{c}}{k}$. We found $h_{c}=0.03$ at $J=0.9$, where the CDW is also destroyed.

The spin gap is defined in a similar way $\Delta _{s}=E_{0}\left ( S_{z}^{tot}=1 \right )-E_{0}\left ( S_{z}^{tot}=0 \right )$ except that $E_{0}\left ( S_{z}^{tot}=-1 \right )$ is not needed due to the spin symmetry. We have calculated the spin gaps for several $J$ with different lattice sizes and the gap is always zero with an error bar in the order of the truncation error. Although numerically we can never rule out a very tiny spin gap in larger sizes, additional evidence to support the vanishing spin gap is found considering the density oscillation period. The effective spin in one unit cell remains a half-integer number, which resembles the spin-half Heisenberg model with gapless spin excitation.

\textit{Correlations.---}
We then turn to the correlation of the localized spins in the CDW. Unlike the charge part, the total spin of the ground state preserves the SU(2) symmetry. The RKKY interaction between the localized spins could lead to possible valence bond solid as it has been reported in the 1D KL at quarter filling\cite{xavier2003dimerization}.

\begin{figure}
\centering
\input{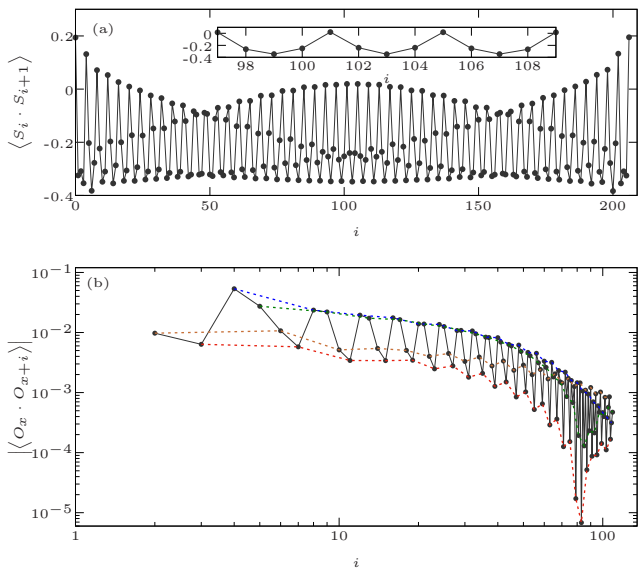}
\caption{\label{Fig6}(Color online) The nearest-neighbor (Fig.\ref{Fig6}$(a)$) and long range dimer-dimer (Fig.\ref{Fig6}$(b)$) correlation of the localized spins at $J=0.9, L=208, n=\frac{3}{4}$. In $(a)$ the average correlation is around -2.0, which is $L$ independent. The incommensurate oscillation refers to the upper bound in the middle with $k=\frac{\pi }{L}$. The amplitude of the incommensurate oscillations has almost no dependence of the lattice length. The inset of $(a)$ shows the dominant oscillation of the correlation with a period of 4, which is the same period as the CDW. In $(b)$ we calculate the correlation at $x=29$ to minimize the boundary effect. The plot uses logarithmic scale on both axis in order to identify the decay mode. The dimer-dimer correlation also has an oscillation period of 4. The 4 color dashed lines are nothing but the correlation of every four points connected, because they each has a different pattern. The correlation functions of the TLL are not shown here as they have monotonic power law decays which agrees with the bosonization theory.}
\end{figure}

We first investigate the nearest neighbor correlation of the localized spins, i.e. $\left \langle S_{i}\cdot S_{i+1} \right \rangle$. Fig.\ref{Fig6}$(a)$ shows that the correlation has $k=\frac{\pi }{2}$ oscillations with a superposition of incommensurate oscillations with $k=\frac{\pi }{L}$ which is size dependent. Generally in a valence bond solid, the dimer order described by the short-ranged spin correlation should be L independent away from the boundary, as the example of the KL on the zigzag ladder at half filling\cite{peschke2018frustrated}. Quantum fluctuations will destroy any incommensurate order in 1D, so the leading order at the weak coupling of $n=\frac{3}{4}$ is just the charge order. The average correlation of the nearest neighbor spins is antiferromagnetic, which is consistent with the RKKY interaction close to half filling. Here the correlation of the localized spins cannot be explained by the low energy effective Heisenberg model with RKKY couplings because of the existence of charge fluctuations. We believe the $k=\frac{\pi }{2}$ oscillation of the localized spin correlation is mainly induced by the $k=\frac{\pi }{2}$ CDW because of the antiferromagnetic coupling between the electron spins and the localized spins. 

For further illustration we study the dimer-dimer correlations which is defined as the two point correlation functions of the localized spins $ \left \langle O_{x}\cdot O_{x+i} \right \rangle $, where $O_{x}=S_{x}\cdot S_{x+1}$ refers to the nearest neighbor spin correlation. The $ \left \langle O_{x}\cdot O_{x+i} \right \rangle $ plotted in Fig.\ref{Fig6}$(b)$ shows a decay over distance, although in an oscillation fashion. The correlation is dominant by the blue line, indicating a slow exponential decay. This is different from the TLL where correlation functions generally has power law decays. It seems that the correlation saturates toward a finite value for the green and red line at around $i=80$, but after analyzing the dimer-dimer correlation at different points, we find that it always 'saturates' near the middle of the lattice. The similar 'saturation' is also found in the TLL phase at $J=2.1,n=\frac{3}{4}$. We argue that this is just an artificial effect due to finite size of system, as the correlation will have a monotonic decay in the limit of an infinite chain. This is consistent with our conclusion that the phase at $n=\frac{3}{4}$ for weak $J$ is dominant by the charge order.

\textit{Summary.---}
We have used the state of the art DMRG method to obtain the ground state of the 1D KL model at $n=\frac{3}{4}$ and provide compelling evidence for CDW in the 1D KL model. Our numerical results is consistent with the bosonization prediction near the phase transition, suggesting that the CDW is driven by the effective Coulomb repulsion, which is qualitatively different from the CDW at higher dimensions\cite{misawa2013charge,peters2013charge}. We also find that CDW is insulating, while the phase at incommensurate filling or generally in TLL is metallic. Under the existence of CDW, the localized spin has formed a similar oscillation pattern, while preserving the total spin SU(2) symmetry. Other magnetic orders like the antiferromagnetic order has not been found in the CDW phase. Our results provide a simple mechanism of the charge order in 1D organic compounds. The emergence of the CDW may have implications for the novel phase diagram of the KL model in three dimensions.

\begin{acknowledgments}
We thank E. Miles Stoudenmire for the helpful discussions on the DMRG implementation. Y.H and C.S.T was supported by the Texas Center for Superconductivity and the Welch Foundation Grant No. E-1146. D.N.S was supported by National Science Foundation Grants PREM DMR-1828019 and by the Princeton MRSEC through the National Science Foundation Grant DMR-1420541. Numerical calculations was completed in part with resources provided by the Center for Advanced Computing and Data Science at the University of Houston.

\end{acknowledgments}

\bibliography{1DKLM}

\end{document}